# Driving behavior-guided battery health monitoring for electric vehicles using machine learning

Nanhua Jiang [a†], Jiawei Zhang[a†], Weiran Jiang[b], Yao Ren[b], Jing Lin[c], Edwin Khoo[c], and Ziyou Song[a*]

[a]*Department of Mechanical Engineering, National University of Singapore, Singapore 117575, Singapore.*

[b]*Farasis Energy USA, Inc., Hayward, CA 94545, USA.*

[c]*Institute for Info COMM Research (I2R), Agency for Science, Technology and Research (A*STAR), 1 Fusionopolis Way, Connexis, Singapore 138632, Singapore.*

## Abstract

An accurate estimation of the state of health (SOH) of batteries is critical to ensuring the safe and reliable operation of electric vehicles (EVs). Feature-based machine learning methods have exhibited enormous potential for rapidly and precisely monitoring battery health status. However, simultaneously using various health indicators (HIs) may weaken estimation performance due to feature redundancy. Furthermore, ignoring real-world driving behaviors can lead to inaccurate estimation results as some features are rarely accessible in practical scenarios. To address these issues, we proposed a feature-based machine learning pipeline for reliable battery health monitoring, enabled by evaluating the acquisition probability of features under real-world driving conditions. We first summarized and analyzed various individual HIs with mechanism-related interpretations, which provide insightful guidance on how these features relate to battery degradation modes. Moreover, all features were carefully evaluated and screened based on estimation accuracy and correlation analysis on three public battery degradation datasets. Finally, the scenario-based feature fusion and acquisition probability-based practicality evaluation method construct a useful tool for feature extraction with consideration of driving behaviors. This work highlights the importance of balancing the performance and practicality of HIs during the development of feature-based battery health monitoring algorithms.

## 1. Introduction

With the rising awareness of the reduction of carbon emissions, electric vehicles (EVs) are established as a leading candidate for the electrification of transportation because of their pollution-free characteristics [1]. Lithium-ion batteries (LIBs) are considered one of the most promising energy storage technologies for EVs, due to their high energy density, long cycle life, and reduction of manufacturing costs [2, 3]. However, the diverse and dynamic operating conditions may aggravate the irreversible side reactions in LIBs, which permanently deteriorate the performance of LIBs and cause capacity and power fade [4, 5], which must be accurately monitored for timely maintenance and replacement [6, 7] to ensure safe and reliable operations

† These authors contributed equally to this work.

* Corresponding authors. E-mail addresses (Ziyou Song): ziyou@nus.edu.sg

# Nomenclature

## Health indicators

| | |
|---|---|
| ACC | Area under charge current |
| ACCCC | Area under constant-current charge current |
| ACCCT | Area under constant-current charge temperature |
| ACCCV | Area under constant-current charge voltage |
| ACCDC | Area under constant-current discharge current |
| ACCDT | Area under constant-current discharge temperature |
| ACCDV | Area under constant-current discharge voltage |
| ACT | Area under charge temperature |
| ACV | Area under charge voltage |
| ACVCC | Area under constant-voltage charge current |
| ACVCT | Area under constant-voltage charge temperature |
| ACVCV | Area under constant-voltage charge voltage |
| ADC | Area under discharge current |
| ADT | Area under discharge temperature |
| ADV | Area under discharge voltage |
| CCCT | Constant-current charge time |
| CCDT | Constant-current discharge time |
| CDE-SOC | Current drop of equal SOC |
| CDET | Current drop of equal time |
| CVCT | Constant-voltage charge time |
| DTA | Differential temperature area |
| DTP | Differential temperature peak |
| DTPL | Differential temperature peak location |
| DTS | Differential temperature slope |
| DVA | Differential voltage area |
| DVS | Differential voltage slope |
| DVV | Differential voltage valley |
| DVVL | Differential voltage valley location |
| ECC | Energy of charge current |
| ECCCC | Energy of constant-current charge current |
| ECCCT | Energy of constant-current charge temperature |
| ECCCV | Energy of constant-current charge voltage |
| ECCDC | Energy of constant-current discharge current |
| ECCDT | Energy of constant-current discharge temperature |
| ECCDV | Energy of constant-current discharge voltage |
| ECT | Energy of charge temperature |
| ECV | Energy of charge voltage |
| ECVCC | Energy of constant-voltage charge current |
| ECVCT | Energy of constant-voltage charge temperature |
| ECVCV | Energy of constant-voltage charge voltage |
| EDC | Energy of discharge current |
| EDT | Energy of discharge temperature |
| EDV | Energy of discharge voltage |
| HCCCT | Highest constant-current charge temperature |
| HCT | Highest charge temperature |
| HCVCT | Highest constant-voltage charge temperature |
| HDT | Highest discharge temperature |
| HT | Highest temperature |
| ICA | Incremental capacity area |
| ICP | Incremental capacity peak |
| ICP-SOC | Incremental capacity peak of SOC |
| ICPL | Incremental capacity peak location |
| ICS | Incremental capacity slope |
| KT | Kurtosis |
| LCCCT | Lowest constant-current charge temperature |
| LCT | Lowest charge temperature |
| LCVCT | Lowest constant-voltage charge temperature |
| LDT | Lowest discharge temperature |
| LT | Lowest temperature |
| MCCCT | Mean constant-current charge temperature |
| MCT | Mean charge temperature |
| MCVCT | Mean constant-voltage charge temperature |
| MDT | Mean discharge temperature |
| MT | Mean temperature |
| RCCCV | Ratio of constant-current to constant-voltage |
| SCC | Slope of charge current |
| SCV | Slope of charge voltage |
| SD | Standard deviation |
| SDT | Slope of discharge temperature |
| SDV | Slope of discharge voltage |
| SE | Sample entropy |
| SK | Skewness |
| TECD | Time of equal current drop |
| TETR | Time of equal temperature rise |
| TEVD | Time of equal voltage drop |
| TEVR | Time of equal voltage rise |
| TRE-SOC | Temperature rise of equal SOC |
| TRET | Temperature rise of equal time |
| VDE-SOC | Voltage drop of equal SOC |
| VDET | Voltage drop of equal time |
| VRE-SOC | Voltage rise of equal SOC |
| VRET | Voltage rise of equal time |

## Other symbols

| | |
|---|---|
| CALCE | Center for advanced life cycle engineering |
| CC | Constant-current |
| CV | Constant-voltage |
| DT | Differential temperature |
| DV | Differential voltage |
| EIS | Electrochemical impedance spectroscopy |
| ELM | Extreme learning machine |
| FRD | Faradic rate degradation |
| GPR | Gaussian process regression |
| HI | Health indicator |
| IC | Incremental capacity |
| ICTR | Increase of charge transfer resistance |
| IMTRAM | Increase of mass transport resistance in active material |
| IMTRE | Increase of mass transport resistance in the electrolyte |
| IOR | Increase of ohmic resistance |
| IQR | Interquartile range |
| LAM | Loss of active material |
| $LAM_{NE}$ | Loss of active negative electrode material |
| $LAM_{PE}$ | Loss of active positive electrode material |
| LCO | Lithium cobalt oxide |
| LFP | Lithium iron phosphate |
| LIBs | Lithium-ion batteries |
| LLI | Loss of lithium inventory |
| NCA | Lithium nickel cobalt aluminum oxide |
| NNs | Neural Networks |
| OCV | Open-circuit voltage |
| PCC | Pearson correlation coefficient |
| RMSE | Root mean square error |
| RVM | Relevance vector machine |
| SOC | State of charge |
| SOH | State of health |
| SVM | Support Vector Machine |
| WOA | Whale optimization algorithm |
| WOA-ELM | Whale optimization algorithm optimized extreme learning machine |

of EVs. State of health (SOH) is an umbrella term referring to any metric that can reflect the capacity and power fade of LIBs. Capacity fade can be directly indicated by the total capacity, while internal resistance is typically used to quantify power capability as it directly restricts the maximum power. Another common metric used in practice for SOH is the discharge capacity at a particular C-rate, which will be affected by both capacity and power fade.

Lab-based measurement and characterization methods mainly focus on experimentally measuring quantities that can directly reflect SOH, such as remaining capacity, internal resistance, or more detailed properties characterizing the internal microstructure changes (e.g., decomposition of active materials and dynamics of stress) [8]. Several mature testing methods can accurately measure battery capacity through coulomb counting [9, 10] in the laboratory. Besides, the internal impedance at various frequencies can also be directly measured via electrochemical impedance spectroscopy (EIS) [11, 12], which is a powerful tool for characterizing and interpreting the impedance-related physical and electrochemical processes inside batteries. Furthermore, the electrode-level structure can also be used as direct proof of the battery degradation, such as loss of active materials and changes in stress. Advanced characterization techniques, including Raman spectroscopy [13], X-ray diffraction [14], and scanning electron microscope [15] have been applied to investigate microscale degradation mechanisms [16]. Louli et al. [17] used operando pressure measurements to detect solid electrolyte interphase (SEI) and provide information about battery health. Nevertheless, these lab-based approaches require either time-consuming tests or complex equipment and thus are not feasible for the on-board battery health diagnosis of EVs in practical scenarios [18, 19].

The limitations of lab-based approaches make on-board SOH estimation necessary. Methods for SOH estimation can be categorized into model-based [20] and data-driven [21]. Model-based methods generally either identify aging-related model parameters from data or even capture the degradation mechanisms of LIBs by modeling how these parameters evolve. Nevertheless, various degradation mechanisms and their interactions remain major challenges for modeling and parameterization as the aging path is complicated and dependent on various operating conditions (e.g., current profile and temperature). In contrast, data-driven methods directly build relations between external measurements (e.g., current, voltage, and temperature) and SOH with sufficient labeled experimental data points, thus avoiding complex modeling of battery dynamics and enabling mechanism-agnostic SOH estimation [22, 23, 24]. With advances in the field of machine learning, data-driven algorithms can handle complex nonlinear problems with higher accuracy and efficiency [25, 26]. Due to these benefits, data-driven models have been extensively studied and applied for battery health diagnosis, such as Gaussian process regression (GPR) [27, 28], relevance vector machines (RVMs) [29], neural networks (NNs) [30] and support vector machines (SVMs) [31]. When applying machine learning models to capacity estimation, it is essential to select appropriate input variables. For practical EVs, time series of current, voltage, and temperature are the major signals available to us for capacity estimation. However, due to a weak direct correlation between these raw signals and SOH, complex machine learning models are usually required at the risk of overfitting.

Instead, health indicators (HIs), which are physics-inspired features derived from the raw signals [32], can increase the reliability of machine learning models and advance possible mechanistic interpretation, thereby drawing increasing attention in recent years. Ramadass et al. [33] developed the constant-current charge time (CCCT) and the constant-voltage charge time (CVCT) as HIs in their first principle-based capacity fade model of LIBs. Since these two HIs are extracted from the entire constant-current (CC) or constant-voltage (CV) phase, they are hardly practical for EV operating conditions. To guarantee high applicability, Liu et al. [34] proposed the time of equal voltage rise (TEVR) as an HI, which can be extracted from only the partial charge process. Similarly, Liu et al. [35] extracted the time of equal voltage drop (TEVD) from the partial discharge process. After selecting an optimal voltage range, the time increment within this range was used as a variable in the capacity degradation model. The basic principle of the above HIs is that the rate of voltage change rises under galvanostatic conditions as internal resistance increases and capacity decreases. Accordingly, the slope of the voltage curve [21] can also be a useful HI. Unfortunately, these time difference-dependent HIs do not generalize across different C-rates. If SOC estimates are available, replacing time windows by SOC windows can alleviate though not eliminate such protocol dependence. Another way around this issue proposed by [36] is to use the current and voltage sequences to calculate discharge power and then construct its integral over time as a novel energy-based HI, which can be applied at different discharge rates. However, the discharge of EVs can be very dynamic and hardly include CC phases unlike the charge process, thus severely limiting the application of HIs that require CC discharge. Deng et al. [37] proposed the temperature-changing rate of an equal charge voltage difference as an HI, representing the amount of irreversible heat during the CC charge process. On top of individual HIs, feature fusion and screening [32, 38, 39] can be used to improve SOH estimation performance by eliminating the less relevant features. In summary, although the new feature-based methods are continuously emerging, the availability of HIs in practical driving scenarios is still not addressed, thereby causing bottlenecks for accurate and reliable SOH estimation in EV applications.

The key to overcoming the above challenge is that the development of a feature-based method should be guided by both performance and practicality. To bridge the aforementioned research gap, this work proposed a feature-based machine learning framework for battery health monitoring, guided by real-world EV driving behaviors. As illustrated in Fig. 1, the core contributions of this work can be summarized in three parts. First, various degradation modes and corresponding HIs were summarized and analyzed from the perspective of degradation mechanisms, which can provide important guidance on how these HIs relate to battery degradation. Second, candidate HIs were evaluated based on both correlation analysis and machine learning model accuracy, using three public battery degradation datasets. Finally, the multi-step screening and scenario-based fusion can build better HIs with higher relevance, and the driving behavior-based feature evaluation method can clearly show the practical acquisition probability of various features, thereby providing useful guidance for enhancing the practicality of HIs. To the best of our knowledge, this is the first work that integrates the evaluations of both performance and quantified practicality with the feature-based machine learning approach for battery health monitoring in EV applications.

## 2. Results and discussion

### 2.1 Analysis of degradation modes

Overall, the battery degradation analysis can be conducted on several levels including systematic degradation effects, scenario-based degradation modes, and fundamental degradation mechanisms, from the external and phenomenological to internal and physical characteristics [40]. The most commonly considered external battery aging characteristics are capacity fade and power fade [4, 5], of which we shall focus on the former in this work as the ground truth of discharge capacity is more readily available in existing datasets. Hereafter, we refer SOH to the ratio of current discharge capacity (at the constant C-rate that the cell is discharged in cycling) to the rated/nominal capacity [41], so a fresh cell should start with SOH close to 100%. EV batteries should be decommissioned from EVs when the SOH reaches 70% or 80% [42, 43] to be recycled [44] or repurposed for second-life applications such as grid storage and household energy systems. Since the battery degradation effects can be attributed to diverse physiochemical processes and their complex interactions, the degradation mode generally works as a bridge between the underlying degradation mechanisms and final degradation effects, which enables a better understanding and analysis of battery aging. The degradation modes of battery electrochemical behaviors can be generally divided into thermodynamic and kinetic ones [4]. The commonly reported three thermodynamic degradation modes are loss of active anode (or negative electrode) material ($LAM_{NE}$), loss of active cathode (or positive electrode) material ($LAM_{PE}$), and loss of lithium inventory (LLI), which would cause the evolution of the open-circuit voltage (OCV) curves of the battery cell. In contrast, the main effect of the kinetic degradation modes is the increase in internal resistance. However, it should be noted that the increase in internal resistance can also reduce the discharge capacity since the cut-off voltage would be reached faster. The kinetic degradation modes can be divided into the increase of ohmic resistance (IOR) and faradic rate degradation (FRD) [45]. While the IOR can be attributed to the increase of the contact resistance and decrease of both electronic and ionic conductivity, the FRD mainly denotes the faradic resistance and capacitance controlled by the charge transfer kinetics and mass transport processes. Based on various processes that control the faradic rate, this work further decoupled FRD into three modes: the increase of charge transfer resistance (ICTR), the increase of mass transport resistance in the electrolyte (IMTRE), and the increase of mass transport resistance in active material (IMTRAM). The introduction of these subdivided degradation modes enables more careful battery aging analysis and possibly more accurate guidance for cell design optimization. Built upon the figures about degradation analysis in [4] and [5], Fig. 2 illustrates the overall relationship between degradation mechanisms, modes, effects, and the various HIs, which will be presented in the following sections.

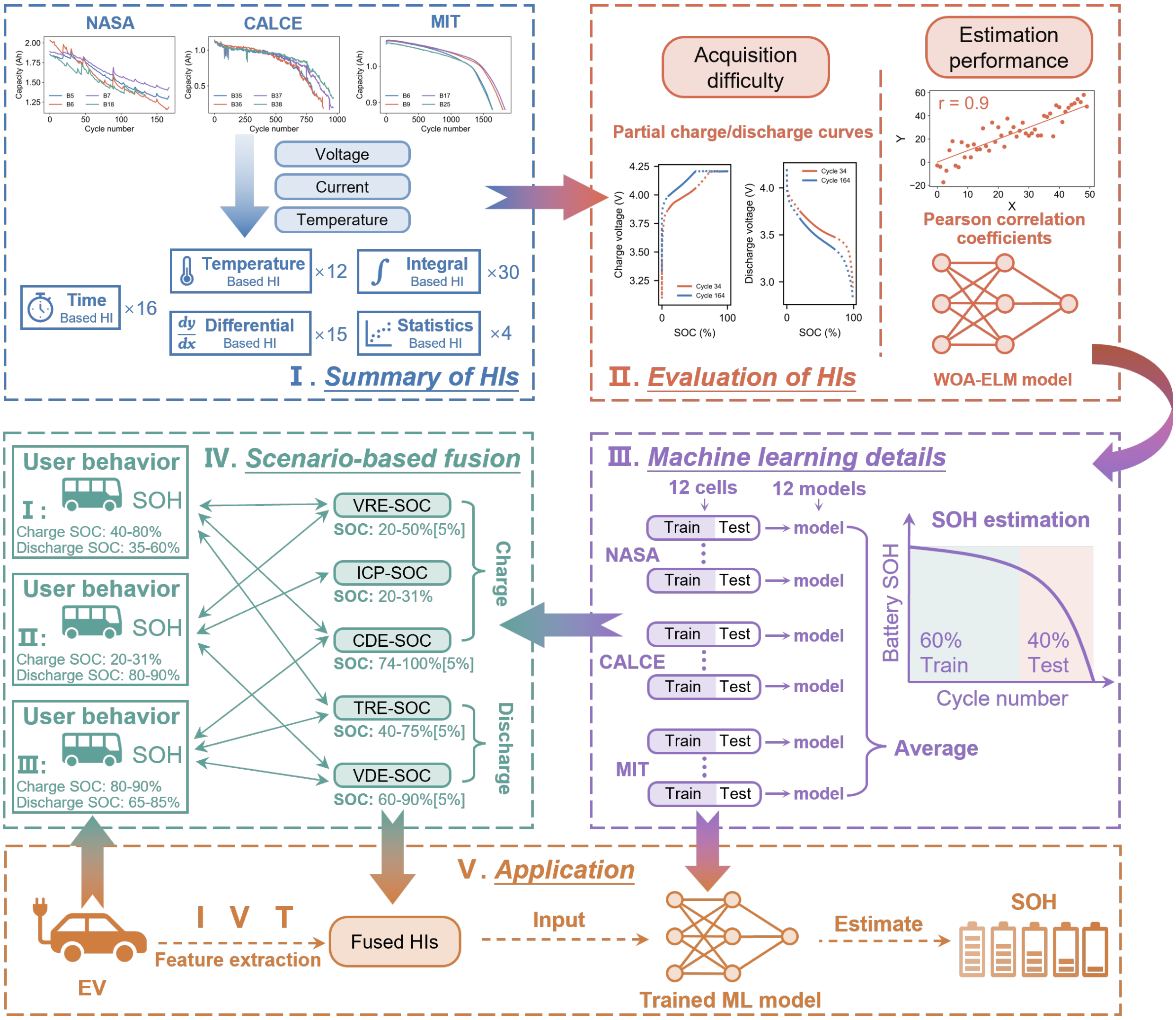

**Fig. 1.** Data-driven battery health diagnosis workflow considering the real-world EV user behavior.

## 2.2 Analysis of various HIs

As summarized in Fig. 2, time-based HIs are those involving time-counting from the related battery measurement signals. One such example is the constant-current discharge time (CCDT) which represents the consumed time of the constant-current discharge process. As internal resistance increases and capacity decreases, the total charge and discharge time will be reduced with constant cut-off voltage, which can be indicated by CCCT [33], CVCT [46, 47], CCDT [48], and time ratio of constant-current to constant-voltage (RCCCV) charging [49]. Since the time-based HIs are phenomenological features for capacity and power fade, they could possibly be attributed to any specific degradation mechanisms, such as the commonly reported SEI growth and lithium plating that would lead to degradation modes of LLI and IMTRE, thereby changing the charge and discharge time. Furthermore, the degradation mode of IMTRAM could especially increase the polarization voltage over some specific SOC intervals, where the slopes of the open-circuit potential (OCP)-stoichiometry curve are large. The reason lies in that the decreased diffusivity in active material particles can lead to a larger lithium concentration gradient and biased surface stoichiometry, thereby leading to significantly biased surface potential over high-slope regions. A possible associated degradation mechanism of IMTRAM can be a central crack of electrode particles, which may increase the tortuosity of the lithium diffusion pathway in the active material, as argued by Barai and Mukherjee [50]. While the above HIs are obtained by counting the total charge or discharge time, other measurements can be added as constraints to construct new HIs which can be obtained with only partial charge or discharge curves (supplemental information).

Temperature-based HIs, derived from battery operation temperatures, can indicate battery health as the heat generation rate rises with increasing internal resistance. The main heat generation sources inside the battery can be attributed to the ohmic heat generation due to the motion of lithium ions, the ohmic contact resistance heat generation between electrodes and current collectors, the reversible heat generation due to the entropy change in the electrodes, and the reaction heat generation due to reaction kinetics [51, 52]. As displayed in Fig. 2, the temperature-based HIs are mainly attributed to the kinetic degradation modes including IOR, ICTR, IMTRE, and IMTRAM. The degradation mechanisms of electrolyte decomposition and loss of electric contact, which induce the degradation mode of IOR, could contribute to the increase in the ohmic heat generation rate. Moreover, the increased reaction heat generation could be attributed to the ICTR, whereas the reversible heat generation could be influenced by the IMTRAM. In summary, the kinetic degradation modes lead to an increase in internal resistance and then change the generation rates of various heat sources. As shown in Supplementary Fig. S1, the temperature profiles from the NASA battery aging dataset [53] demonstrate that the heating rate during both CC charge and discharge rises with increasing cycle numbers. Therefore, characteristic temperature values such as the highest, mean, and lowest temperatures of various operating scenarios are capable of effectively reflecting the degradation states of batteries [53]. It should be noted that these statistical temperature values will heavily depend on the ambient environment and any implemented battery cooling scheme.

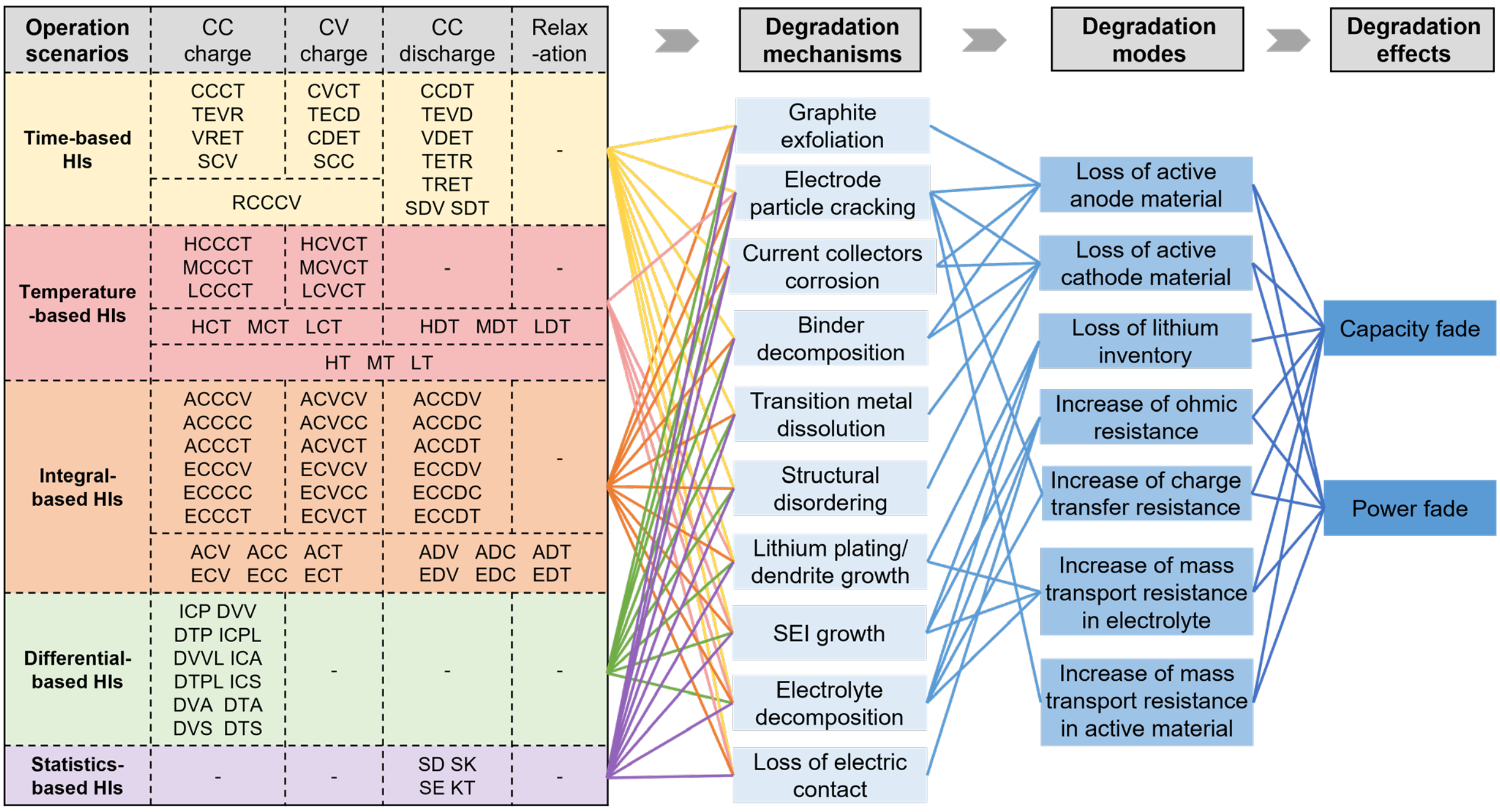

**Fig. 2.** Summary of HIs and possibly associated degradation mechanisms, modes, and effects [4][5].

Integral-based HIs are calculated by integrating the measurable variables over time. They are classified into two categories including area-based and energy-based. While the former denotes the area under the measurement signals [54, 55], which is no other than the integral of the signal over time (e.g., the area under the CC charge current (ACCCC) [51, 53]), the latter is calculated by integrating the signal square over time, which originates from the definition of the 'energy' of the signal, as defined below in Eq. (1):

$$E = \int_{-\infty}^{\infty} [X(t)]^2 dt \tag{1}$$

where $X(t)$ represents the measurable signal, and $t$ is the time. It should be noted that the actual energy-based HIs are always integrated over a finite time range. The energy-based HIs are expected to perform well under dynamic driving conditions since they can reflect the throughput of either capacity or energy, which is inaccessible for area-based HIs that would cancel charge and discharge capacity. Similar to the time-based HIs, the integral-based HIs can be attributed to both thermodynamic and kinetic degradation modes. As shown in Fig. 2, the energy-based HIs about voltage [55], current, or temperature can be flexibly extracted during different operation phases and show satisfactory performance [53, 56, 57]. However, they should be carefully used as they tend to accumulate errors in the signal over a prolonged period of time.

Differential-based HIs employ differential analysis methods, encompassing IC, DV[39], and differential temperature (DT) to estimate battery SOH. The values of the IC peak (ICP) [58], the IC peak location (ICPL) [59], the IC peak area (ICA) [60], and the IC peak slope (ICS) [53] can be extracted for battery capacity degradation mechanism identification, such as graphite exfoliation due to the co-intercalation of solvent molecules [5], the irreversible structural deformation loss of metal oxide cathode materials caused by Jahn-Teller distortion, transition metal dissolution [61] and structural disordering [62]. These aging mechanisms would cause the distortion of the open-circuit potential curves of both electrodes, thereby changing the differential-based features. Although DT analysis is also used to analyze the temperature curve [61], fewer degradation mechanisms can be interpreted compared with temperature-based HIs. It should be noted that differential-based HIs only work effectively during low-rate applications, as the polarization induced by a high C-rate could obfuscate the thermodynamic behavior of the voltage. Another issue is that the peaks and valley features are susceptible to measurement noises due to the noise amplification property of differential analysis, therefore the filtering algorithms are used in this work for subsequent analysis and calculation.

Statistics-based HIs are derived from time series data by treating segments of the series as samples of a random variable and computing relevant statistical measures. Although these statistics are usually not very interpretable regarding the physics of battery degradation, they are worth investigating owing to their conceptual simplicity. More details about the specific definitions of the four statistics-based HIs can be found in the supplemental information.

## 2.3 Correlation analysis and estimation performance of HIs

After comprehensively reviewing the literature, we summarized 77 candidate HIs that are extracted from either partial or full battery operation curves, where "P" denotes partial and "F" represents full, as given in Table 1. It should be noted that although we remained the feature extraction methods in the literature, such as extracting some HIs from full curves, they can in principle also be obtained from partial curves. This work focused on reviewing and reproducing the existing features and extraction methods, while the further investigation about the effects of the completeness of operation curves would be done in future work. To evaluate the correlation and estimation accuracy of HIs, three public battery aging datasets (NASA, CALCE, and MIT) are used, where each battery cell is cycled under the same protocol, but different cells may be cycled under different protocols. The specifications of the three battery degradation datasets and the selected 12 cells are listed in the supplemental information.

The correlations between HIs and SOH determine the significance of various HIs in reflecting the actual battery state, guiding the selection of the most relevant HIs for further evaluation. They can be quantified by one of the many correlation coefficients [63], which statistically measure the strength of the dependence between two variables. Various commonly used options include Pearson correlation coefficient (PCC) [64], Spearman correlation coefficient [65], Kendall correlation coefficient [66], and Grey relation analysis [59]. Since PCC is most widely used for the correlation between variables, it is adopted in this work to analyze the correlation between various HIs and SOH. The PCC is calculated as follows:

$$r = \frac{\sum_{i=1}^{c}(x_i - \overline{x})(y_i - \overline{y})}{\sqrt{\sum_{i=1}^{c}(x_i - \overline{x})^2}\sqrt{\sum_{i=1}^{c}(y_i - \overline{y})^2}} \tag{2}$$

where the sum is over cycle 1 to $c$, while $\overline{x}$ and $\overline{y}$ represent the mean of the HI and of SOH, respectively. $x_i$ and $y_i$ represent the HI and SOH at cycle $i$, respectively. The value of the correlation coefficient is between -1 and +1, with a larger absolute value indicating stronger dependence, although PCC only characterizes linear correlation. For each HI to be evaluated, the PCC is first calculated for the 12 cells separately before being averaged as the evaluation index.

Since the intervals of the partial-curve-based HIs can affect the performance, the grid search method is used to optimize the intervals over several combinations of starting and ending points for maximizing the absolute PCCs (instead of the machine learning model performance, the PCC is selected as the optimization objective for higher computational efficiency). The setting of the grid search method and the result of optimal intervals are referred to Supplementary Table S5. Further performance evaluation and comparison are conducted by using the optimal interval of each HI, where the absolute PCC values are listed in Table 1.

The best-performed time-based HI is the CCDT, which reached an absolute PCC value of 0.99722 showing a very high correlation with SOH. Although the CVCT performed worst among the time-based HIs, it still reached a relatively high absolute PCC value of 0.86933. However, the PCC disparity between the best and worst temperature-based HIs, namely HDT and LT, is very large (i.e., 0.76011 and 0.14345, respectively). The major reason is that the lowest temperature is generally determined by the ambient temperature, which is essentially independent of the increased heat generation rate with degradation. A similar situation has arisen among the integral-based HIs, that is the ADC reached the highest PCC value of 0.99846 among all 77 candidate HIs, whereas the EDT only has 0.53038. For differential-based HIs, the incremental capacity-based HIs are in general better than the differential voltage-based HIs. The possible reason lies in that the variation of IC-based HIs at different states of aging is more obvious for reflecting the degradation mechanisms of LLI and LAM than DV-based HIs. Most statistics-based HIs show poor performance except the SE, which reached an absolute PCC value of 0.88828.

To further compare different categories of HIs, a statistical box plot is used to summarize the overall evaluation results. As can be seen in Fig. 3, the absolute PCC results of five categories of HIs are analyzed by calculating several typical statistical values. The box in light blue represents the lower and upper quartiles of a set of data, and the interval between their values is called the interquartile range (IQR). Here the boundaries of the whiskers are set as the upper quartile plus or the lower quartile minus 1.5 times the IQR, where the points outside the whisker boundaries would be marked as outliers with pink dots.

Moreover, the mean and median of a set of PCC values would be represented by the green triangle and red line. The best overall performer is time-based HIs, whose mean and median are both greater than 0.95, and the IQR is quite small. Similar to the time-based HIs, since most integral-based HIs can directly indicate the charge or discharge capacity, it shows the second-best performance. However, it has more outliers such as EDT and ECT, which contain a significant amount of redundant information that can weaken the representation ability of the feature. The differential-based and statistics-based HIs come next, with mean PCCs around 0.75 and 0.77. It is seen that the PCCs of most lowest-temperature-based HIs are not high. The major reason for this lies in that the minimum temperature would be largely affected by the ambient temperature, instead of the increment of heat generation rates due to battery degradation.

Besides employing correlation analysis, the RMSEs of SOH estimation using machine learning algorithms (ELM and WOA-ELM) are used for the performance evaluation of HIs in this work. Specifically, various HIs are applied as input features, while the battery SOH is set as the output of the machine learning models, which are kept identical across all HIs to ensure a fair comparison. After the HIs are extracted, they can be fed to a machine learning model to estimate SOH. Since a typical cell has only hundreds of to over a thousand cycles, the size of a training set will be small so it is better to use a simple model to avoid overfitting. In this work, the extreme learning machine (ELM) [67], which is a two-layer neural network with randomly generated first-layer parameters, is used for SOH estimation due to its fast and simple training process as described in the following:

$$H = g(WX + b1_m^T), \qquad \beta = YH^+ = YH^T(HH^T)^{-1}, \qquad \hat{Y} = \beta H, \tag{3}$$

where $X \in \mathbb{R}^{d_i \times m}$ is the input matrix containing $m$ data points, each of dimension $d_i$, while $H \in \mathbb{R}^{d_h \times m}$ and $Y \in \mathbb{R}^{1 \times m}$ (SOH output is one-dimensional) contain the $m$ instances of hidden layer output and final output, respectively. Here the entries of the weight matrix $W \in \mathbb{R}^{d_h \times d_i}$ and bias vector $b \in \mathbb{R}^{d_h}$ for the first layer are randomly generated, and $1_m^T$ is an all-one $m$-vector to broadcast $b$ to all $m$ instances. Hence, the first layer simply randomly projects the input to $d_h$ different one-dimensional subspaces, before the sigmoid activation function $g(\cdot)$ acts on each projection outcome to obtain $H$. Given the first-layer parameters, the second layer reduces to using linear regression to map these random features in $H$ to output. Therefore, the linear coefficients $\beta \in \mathbb{R}^{1 \times d_h}$ can be determined in closed form as $\beta = YH^+$ using the psuedo-inverse $H^+ \in \mathbb{R}^{m \times d_h}$ and the output data $Y$. Finally, the output of the fitted model can be calculated as $\hat{Y} = \beta H$. In summary, the ELM is essentially generating features from input by random projection and perform linear regression on these random features. Since no nonlinear optimization is required, training an ELM is much faster than optimizing all the parameters in a two-layer network, even if the former uses several times more hidden nodes, while the former can achieve comparable accuracy to the latter and might even be less susceptible to overfitting [67].

To compare with ELM, we also use a fully-optimized two-layer network of the same architecture as a second model. To take advantage of the two-layer architecture as the ELM does, we use a gradient-free meta-heuristic optimization algorithm to optimize the first-layer parameters only, while treating the second layer ones as determined by Eq. (3). Meta-heuristic optimization algorithms have been widely studied and used due to their potential to avoid the local optima, compared to gradient-based methods. The previous research [68] has shown that the prediction performance of the whale optimization algorithm-optimized ELM (WOA-ELM) is better than ELM, genetic algorithm-optimized ELM, cuckoo search-optimized ELM, and dandelion algorithm-optimized ELM. Hence, the WOA-ELM is used as the second machine learning model for HI evaluation.

The whale optimization algorithm (WOA) [69] is a swarm optimization algorithm inspired by the humpback whales' special hunting strategy:

$$X_{t+1}^{i} = \begin{cases} X_t^{ref} - ar_1 \odot \left| r_2 \odot X_t^{ref} - X_t^{i} \right|, & \text{if } p < 0.5, \\ X_t^{*} + \left| X_t^{*} - X_t^{i} \right| \odot e^{bL} \odot \cos(2\pi L), & \text{if } p \geq 0.5, \end{cases} \tag{4}$$

$$X_t^{*} = \text{best from } X_k^{i}: 1 \leq i \leq s, 1 \leq k \leq t, \quad X_t^{ref} = \begin{cases} \text{randomly chosen from } X_t^{i}: 1 \leq i \leq s, & if\ a \geq 1, \\ X_t^{*}, & if\ a < 1, \end{cases} \tag{5}$$

where $X_{t+1}^{i} \in \mathbb{R}^{d}$ denotes member $i$ of the $s$-size population at iteration $t$. Here $r_1 \sim U^{d}([-1,1])$ is a random vector of length $d$ whose entries are independent and uniformly distributed over $[-1,1]$. Similarly, $r_2 \sim U^{d}([0,2])$ and $L \sim U^{d}([-1,1])$, while $p \sim U([0,1])$. These random variables are drawn independently for each member $i$ at each iteration $t$. The "$\odot$" operator denotes element-wise multiplication between vectors. Besides, $b = 1$ is a constant, while $a$ is scheduled to decrease linearly from 2 to 0 across a specified number of iterations to control the exploration-exploitation trade-off for global optimization.

A humpback whale recognizes its prey by cooperating with the group and creates bubble nets to encircle it when hunting. As shown in Eq. (4) and Eq. (5), whales are initially (when $a \geq 1$) pushed away from each other to explore globally, but are attracted to the current optimizer in a later stage (when $a < 1$). Half of the times, the whales also do a spiral motion towards or away from the current optimizer, as shown by Eq. (4). In WOA-ELM, the above algorithm is applied to optimizing the first-layer parameters of a two-layer network.

In this work, the number of hidden neurons in both ELM and WOA-ELM is set to $d_h = 20$. The maximum number of iterations and population size for WOA are set to $t_{\max} = 30$ and $s = 20$, respectively. To evaluate the HIs in terms of estimation performance, a model is trained and tested for each HI and for each of the 12 cells. For a particular HI, its testing errors of the 12 models are averaged as the final evaluation index. In the cycling data of each cell, the first 60% of the whole life cycles are used for training and the remaining 40% for testing. This train-test split can better examine the generalizability of the HI and model across cycles at different stages of aging.

As can be demonstrated in Table 1, the evaluation indices include the absolute PCC value, the SOH estimation RMSE of both ELM and WOA-ELM. In each category, the HIs were sorted in descending order by the absolute PCC due to its deterministic algorithm, instead of the estimation error of the machine learning models, whose parameters are in general initialized or trained with a certain degree of randomness. The higher PCC value generally corresponds to the lower estimation RMSE. It can also be seen that almost all the RMSE results of the WOA-ELM are less than the ELM (some individuals are equal to the ELM), which fully verifies the superiority of the WOA-ELM model benefiting from the optimization of weights and threshold settings. It is also worth noting that there is a mismatch between the performance ranking of PCC and machine learning models, where some HIs with very low correlation have decent estimation performance. The main reason for this conflict is that the PCC can only measure the linear dependence between two variables, whereas the relationship between HIs and SOH can be highly nonlinear. In future work, the correlation metrics that can measure nonlinear dependence can be used for feature evaluation.

**Table 1:** The evaluation results of all candidate HIs based on correlation analysis and machine learning model performance.

| | | RMSE | | | | | RMSE | | |
|---|---|---|---|---|---|---|---|---|---|
| **HI** | **Abs.PCC** | **ELM** | **WOA-ELM** | **Curves*** | **HI** | **Abs.PCC** | **ELM** | **WOA-ELM** | **Curves*** |
| **Time-based HIs** | | | | | ACCCV | 0.99278 | 1.45% | 0.64% | F |
| CCDT | 0.99722 | 0.25% | 0.21% | F | ACCCC | 0.99240 | 1.68% | 0.67% | F |
| TEVD | 0.99477 | 0.58% | 0.52% | P | ECCCC | 0.99239 | 1.56% | 0.69% | F |
| SDV | 0.99321 | 0.63% | 0.60% | P | ECCDV | 0.98583 | 0.53% | 0.49% | F |
| CCCT | 0.99243 | 1.12% | 0.67% | F | ACCDV | 0.98048 | 0.51% | 0.45% | F |
| SCV | 0.98683 | 1.06% | 0.88% | P | ACCCT | 0.96089 | 1.44% | 0.97% | F |
| VDET | 0.97846 | 1.79% | 0.84% | P | EDV | 0.95798 | 1.36% | 1.19% | F |
| TEVR | 0.96255 | 1.35% | 0.97% | P | ADV | 0.95446 | 0.37% | 0.33% | F |
| VRET | 0.95801 | 2.08% | 1.40% | P | ECCCT | 0.89206 | 1.99% | 1.19% | F |
| CDET | 0.95409 | 1.47% | 1.28% | P | ACVCT | 0.85080 | 11.54% | 2.27% | F |
| TETR | 0.95233 | 1.13% | 1.06% | P | ECVCT | 0.83963 | 22.48% | 2.59% | F |
| SDT | 0.94633 | 1.11% | 0.96% | P | ACCDT | 0.82604 | 4.70% | 0.89% | F |
| SCC | 0.94343 | 2.44% | 2.33% | P | ADT | 0.80190 | 5.22% | 1.46% | F |
| TECD | 0.94216 | 2.50% | 2.31% | P | ECVCV | 0.79709 | 5.48% | 2.82% | F |
| TRET | 0.93470 | 1.16% | 1.11% | P | ACVCV | 0.79699 | 5.78% | 2.65% | F |
| RCCVV | 0.91920 | 6.41% | 1.87% | F | ACVCC | 0.79629 | 3.77% | 2.53% | F |
| CVCT | 0.86933 | 8.36% | 2.63% | F | ECV | 0.70941 | 11.06% | 2.43% | F |
| **Temperature-based HIs** | | | | | ACV | 0.68663 | 5.54% | 2.68% | F |
| HDT | 0.76011 | 2.64% | 2.54% | F | ECVCC | 0.68070 | 4.17% | 2.73% | F |
| HT | 0.76010 | 2.63% | 2.33% | F | ACT | 0.65393 | 21.77% | 2.38% | F |
| MDT | 0.54743 | 2.15% | 2.07% | F | ECCDT | 0.62520 | 1.87% | 1.38% | F |
| MT | 0.35116 | 379.98% | 2.21% | F | ECT | 0.62237 | 17.06% | 1.93% | F |
| HCCCT | 0.28391 | 96.13% | 2.20% | F | EDT | 0.53038 | 1.99% | 1.91% | F |
| MCT | 0.26533 | 87.55% | 2.21% | F | **Differential-based HIs** | | | | |
| LCCCT | 0.24425 | 162.88% | 1.71% | F | ICA | 0.98549 | 7.39% | 0.83% | F |
| HCVCT | 0.24182 | 8.96% | 1.93% | F | ICP | 0.98108 | 1.49% | 0.84% | P |
| HCT | 0.23012 | 30.56% | 1.96% | F | DVV | 0.97890 | 10.89% | 1.26% | P |
| MCVCT | 0.22980 | 2.58% | 2.26% | F | ICPL | 0.90153 | 12.89% | 2.06% | P |
| MCCCT | 0.22897 | 99.36% | 1.88% | F | ICS | 0.86378 | 3.56% | 2.28% | P |
| LCVCT | 0.22383 | 2.78% | 2.56% | F | DVA | 0.86206 | 2.50% | 2.42% | F |
| LDT | 0.16122 | 333.51% | 1.98% | F | DTA | 0.80345 | 1.96% | 1.73% | F |
| LCT | 0.14976 | 346.66% | 71.34% | F | DVS | 0.69462 | 12.00% | 2.96% | P |
| LT | 0.14345 | 789.64% | 11.43% | F | DVVL | 0.58735 | 30.59% | 2.49% | P |
| **Integral-based HIs** | | | | | DTS | 0.44465 | 6.30% | 2.12% | P |
| ADC | 0.99846 | 0.27% | 0.23% | F | DTP | 0.43407 | 4.46% | 2.11% | P |
| ECCDC | 0.99826 | 0.22% | 0.19% | F | DTPL | 0.40335 | 1.78% | 1.14% | P |
| EDC | 0.99775 | 0.32% | 0.29% | F | **Statistics-based HIs** | | | | |
| ACCDC | 0.99644 | 0.27% | 0.24% | F | SE | 0.88828 | 1.81% | 1.57% | F |
| ECC | 0.99597 | 0.53% | 0.41% | F | KT | 0.81226 | 5.25% | 2.23% | F |
| ACC | 0.99363 | 0.60% | 0.57% | F | SK | 0.78868 | 2.96% | 2.40% | F |
| ECCCV | 0.99301 | 0.78% | 0.64% | F | SD | 0.61046 | 10.25% | 2.98% | F |

*If the HI is extracted from the partial or full operation curves? "P" denotes partial, "F" denotes full.

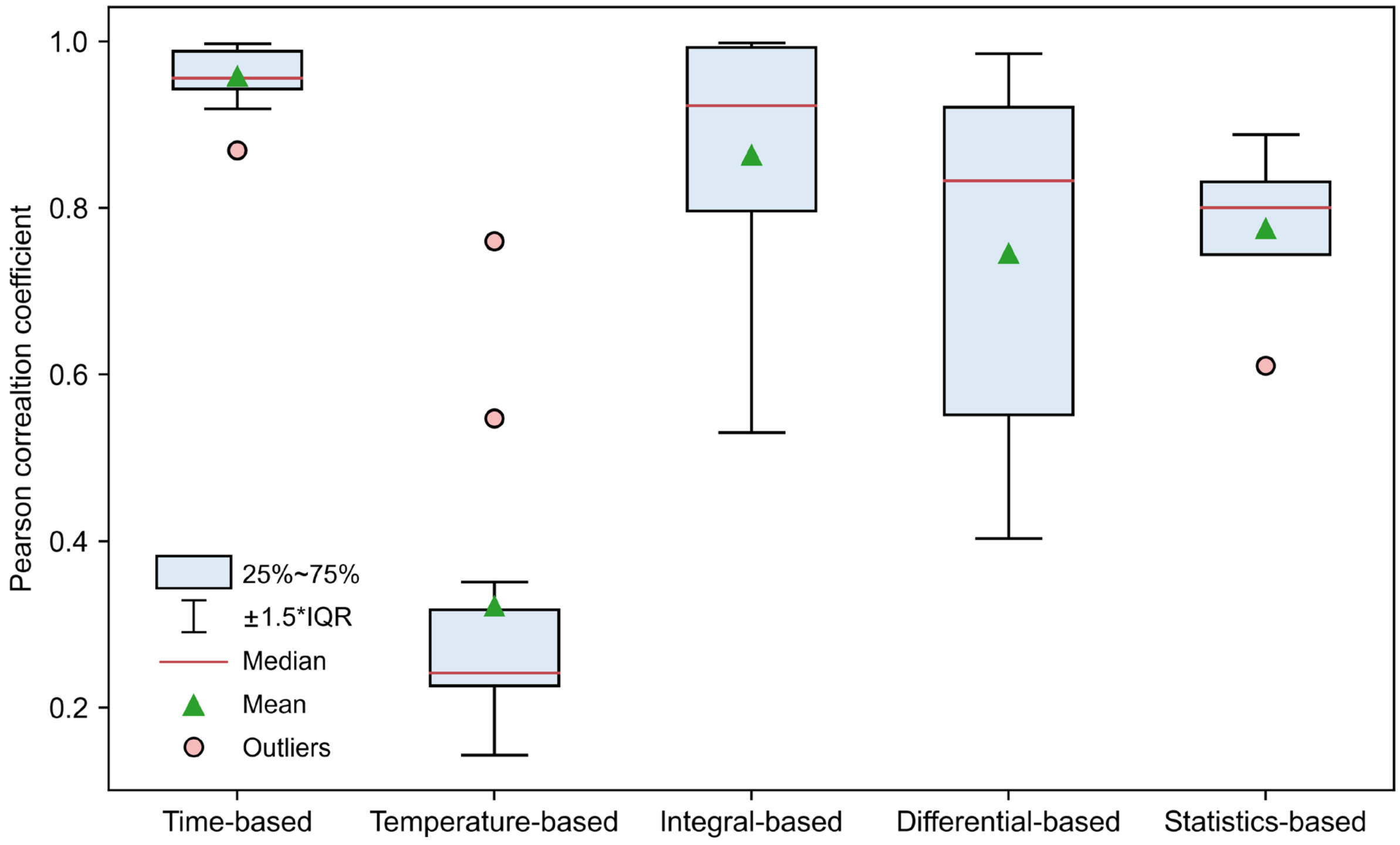

**Fig. 3.** The statistical PCC results of five categories of HIs.

## 2.4 Multi-step screening workflow

Feature screening is generally an important step in feature engineering, especially when lots of candidate features are available. In most previous studies, only correlation coefficients and estimation performance were used as evaluation indices in feature engineering for data-driven battery SOH estimation tasks. However, some HIs can only be extracted when full charge or discharge curves are available, which is not realistic in many practical scenarios (e.g., EVs). Furthermore, although some features can be obtained from partial operation curves, they might be difficult to acquire due to the small occurrence probability of the associated SOC intervals (e.g., extremely low SOC). Therefore, acquisition probability is an important feature evaluation index reflecting the practicality of features. To this end, we define the acquisition probability of HIs as the number of successful extractions with SOC interval divided by the total number of driving samples, as SOC is a typical feature recorded in EV data. Specifically, a large-scale publicly available EV driving dataset that was collected from more than 3 million Chinese EVs, containing real-world operating states and vehicle characteristics [70], is used to evaluate the practicality of HIs in this work (The overall description for the large-scale publicly available EV driving dataset is displayed in Supplementary Table S1). It should be noted that some HIs (e.g., time-based HIs) will be converted to SOC-based HIs for probability calculation in the following sections. In this EV dataset, more than 2 million pieces of data about the real-world states of vehicles during a driving or charge process are provided, including starting SOC and ending SOC with a resolution of 1%. Considering the appearance of each 1% SOC interval as an independent event, the histogram of appearance amount distribution of various SOC intervals demonstrated that the SOC intervals from 35% to 95% cover approximately 82% of total SOC usage, as shown in Supplementary Fig. S2 [70].

Enabled by the multi-dimensional evaluation of 77 candidate HIs, a systematic multi-step screening workflow is proposed to screen out superior HIs with both good performance and practicality, as illustrated in Fig. 4. The first step is based on the PCC evaluation results, where 43 candidates with absolute PCC less than 0.9 were eliminated to ensure high correlations with battery SOH. Next, the remaining 34 HIs were further filtered based on the criteria of whether the HIs can be extracted from the partial charge or discharge curves, which is shown in Table 1. As a result, only twelve time-based and three differential-based HIs remained. The third step would make trade-offs among the HIs that have similar principles to avoid information redundancy. For example, TEVD, VDET, and SDV all reflect the change rate of discharge voltage, which reveals that only one of them should be retained. Since the extraction reference of the VDET is the time interval, which can be converted to SOC interval for the calculation of acquisition probability, it finally remained. Another example is the ICP and ICPL, which are both extracted from the IC curve. The ICPL is screened out due to higher PCC and higher estimation RMSE. The final step would take into account both acquisition probability and estimation performance. To calculate the acquisition probability, the HIs would be first converted to SOC-based features that use the SOC interval as a reference for extraction (e.g., from VRET to VRE-SOC). Although the specific value of the HIs in CV-charge stage might be scaled by this conversion, the variation trend of the HIs with degradation will not be much affected (Supplementary Fig. S3). With consideration of both performance and practicality, five SOC-based HIs were finally selected, as listed in Table 2.

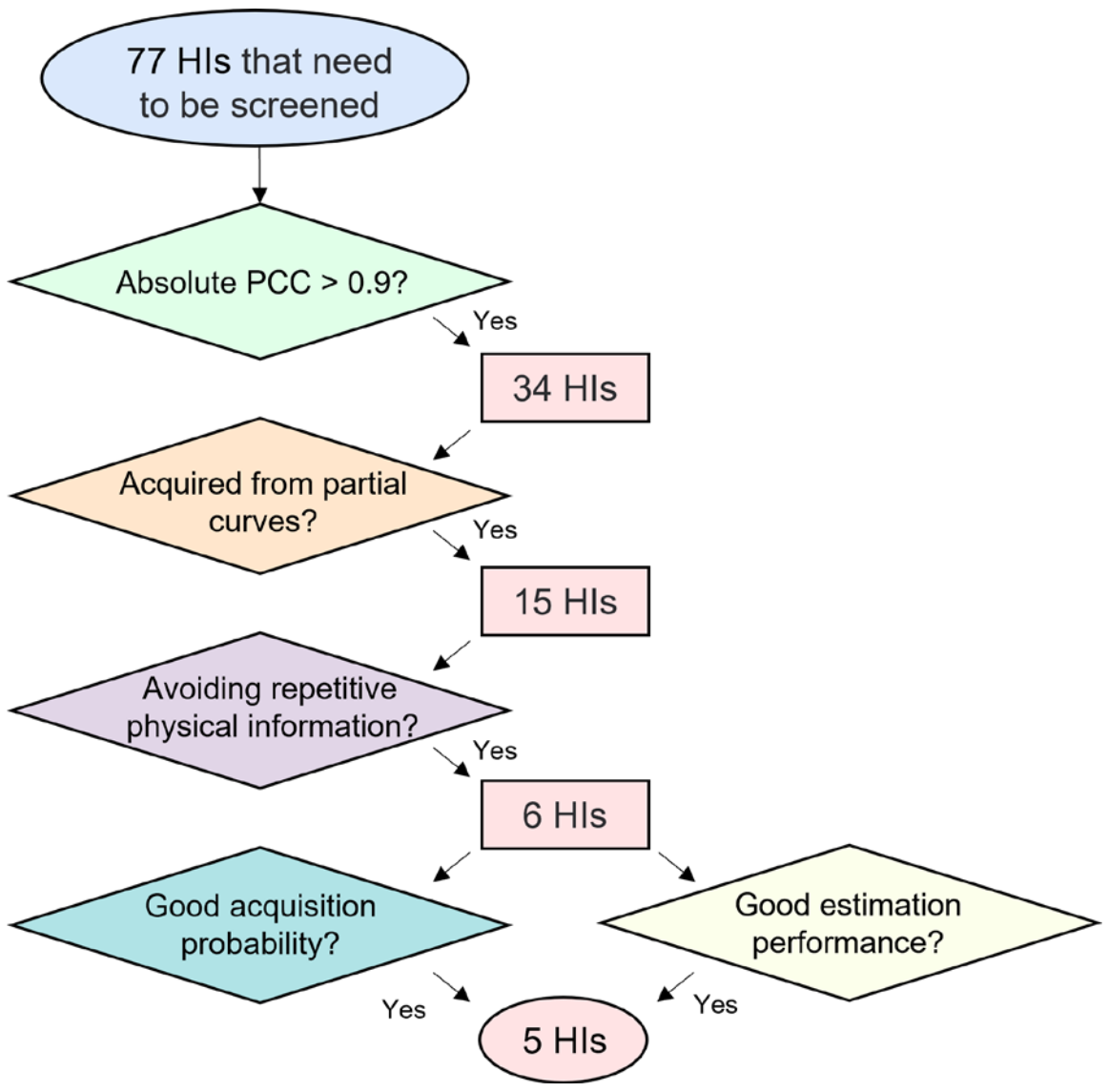

**Fig. 4.** Multi-step screening workflow.

**Table 2:** Screened five SOC-based HIs.

| HI | Definition |
|---|---|
| VRE-SOC | Charge voltage rise of equal SOC interval |
| ICP-SOC | IC peak versus SOC |
| CDE-SOC | Charge current drop of equal SOC interval |
| TRE-SOC | Discharge temperature rise of equal SOC interval |
| VDE-SOC | Discharge voltage drop of equal SOC interval |

To further analyze the variation trends of five SOC-based HIs with increasing battery degradation, the associated operation profiles from the NASA degradation dataset are plotted in Fig. 5. Specifically, the charge voltage, charge current, discharge voltage, and discharge temperature profiles are plotted in Fig. 5(a), (c), (d), and (e). The variation trends of these measurements are clear and directly reflect the degradation effects of capacity and power fade. The IC curves calculated from the original charge voltage data are shown in Fig. 5 (b), where the IC peak moves towards both lower magnitude and lower SOC as the degradation worsens, implying the shrinking process of the OCP curve due to LLI and LAM. Therefore, the SOC interval required to extract the ICP-SOC across the whole life cycle can be determined by the peak locations of the first and last cycles.

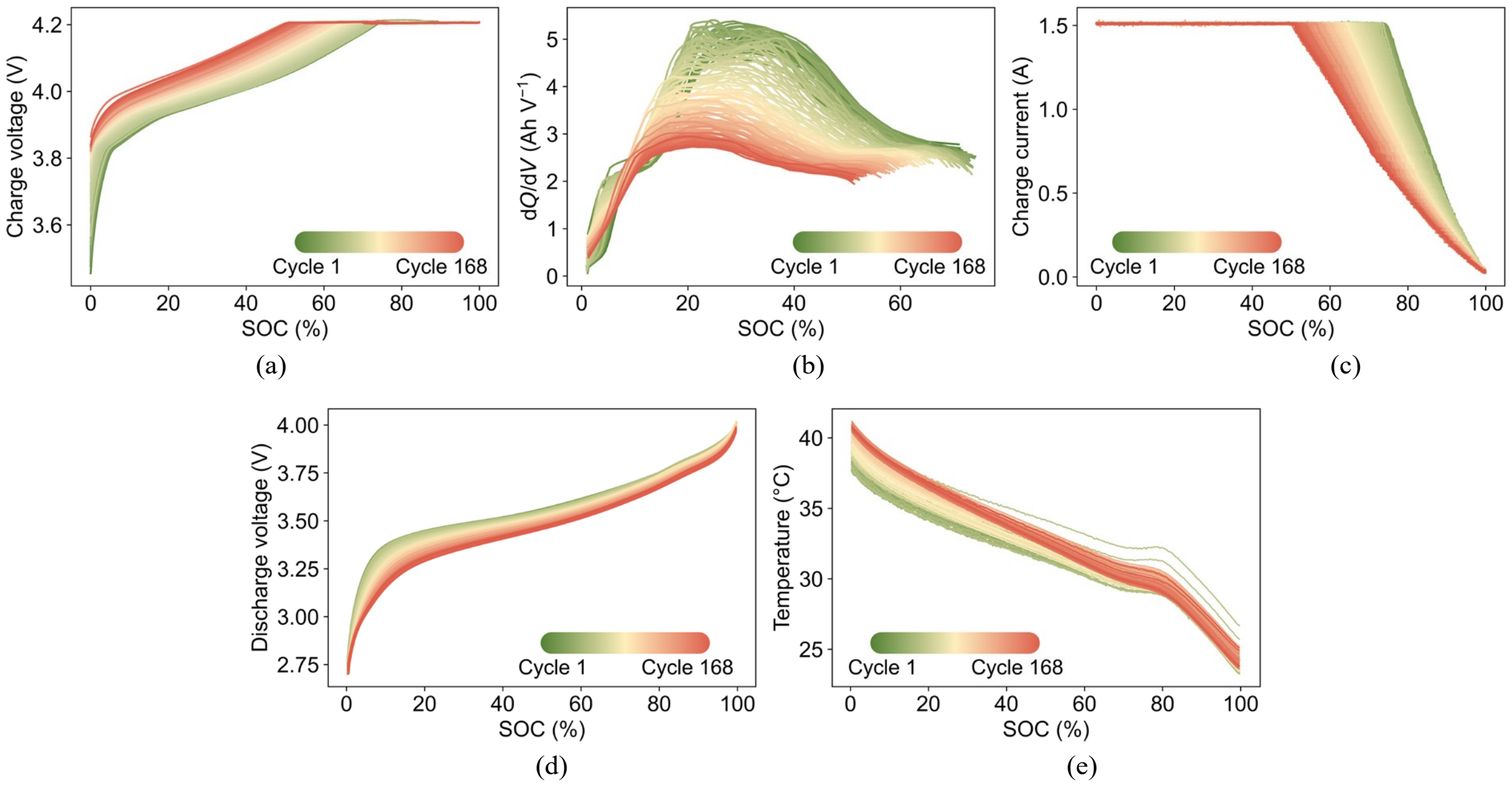

**Fig. 5.** Battery performance degradation data from NASA dataset for extracting SOC-based HIs: (a) CC-CV charge voltage curves for extracting VRE-SOC, (b) CC charge IC curves for extracting ICP-SOC, (c) CC-CV charge current curves for extracting CDE-SOC, (d) CC discharge voltage curves for extracting VDE-SOC, and (e) CC discharge temperature curves for extracting TRE-SOC. While the current, voltage, and temperature profiles are directly obtained from original dataset, the IC curve is calculated from the voltage curves.

Although the SOC interval of the ICP-SOC can be determined, the other four SOC-based HIs have no fixed SOC interval reference used for feature extraction. Therefore, it is worth investigating how the estimation performance is affected by different SOC intervals. Here, the relations between the SOC intervals and absolute PCCs of four SOC-based HIs are shown in Fig. 6 in the form of a heat map. As can be seen in Fig. 6(a), the PCCs of VRE-SOC from 20% SOC to 50% SOC are all greater than 0.9, showing very high correlations with battery SOH. Furthermore, any 5% SOC interval in this range is sufficient to reach a high PCC value (i.e., 20% to 25% and 25% to 30% showed similar performance). However, some SOC intervals outside this range showed relatively poor performance, which implies that the variation of features with degradation could be irregular, thereby decreasing the overall correlations with SOH. The results of the CDE-SOC are shown in Fig. 6(b), which clearly showed that several SOC intervals have very low PCCs, such as the interval from 60% to 90% SOC. This is because both CC and CV charge modes could be contained for most early cycles when the SOC interval is too large. As a result, the variation trend of the current drop shows obvious non-monotonicity (i.e., first increased and then decreased), which is however inconsistent with the pattern of capacity decay and then leads to low PCC. This is also verified by the high PCCs of all possible SOC intervals between 75% and 100%, which only contain CV charge mode. It can be seen in Fig. 6(c) that most SOC intervals of TRE-SOC showed satisfied performance except for the interval from 65% SOC to 70% SOC with PCC of 0.61, mainly because the increase in heat generation rate with degradation in this interval is quite slight, as shown in Fig. 5(e). The heat map can also indicate the best SOC intervals for feature extraction. For example, it can be suggested from Fig. 6(d) that the optimal range for extracting the VDE-SOC is from 10% to 40% SOC, as any interval of 5% can reach a PCC higher than 0.9.

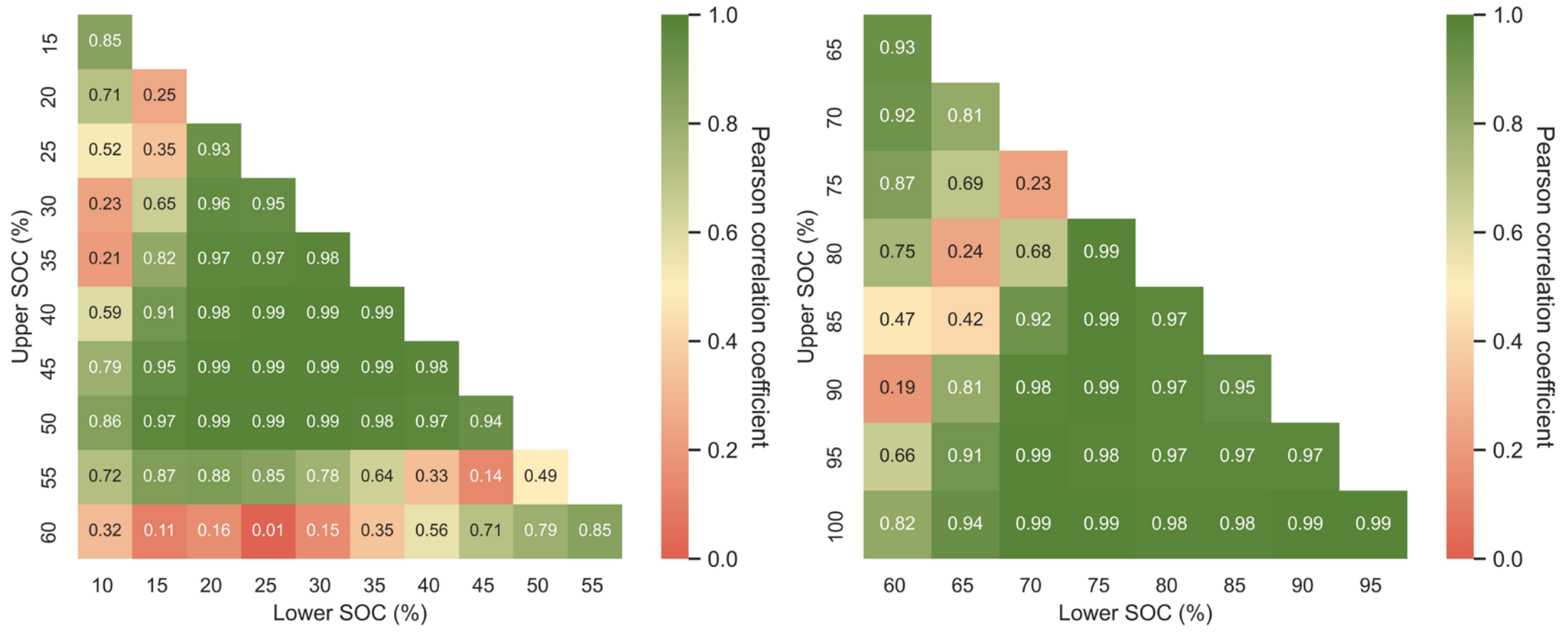

(a) VRE-SOC (b) CDE-SOC

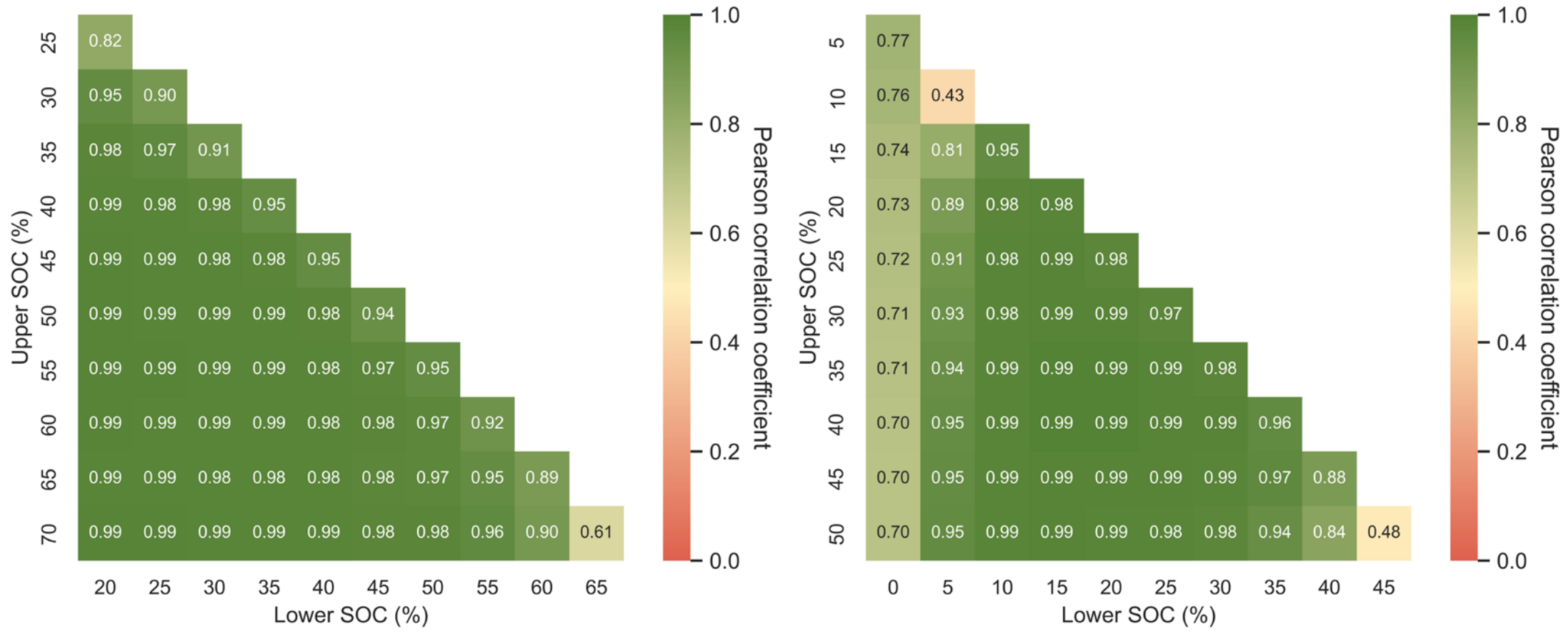

(c) TRE-SOC (d) VDE-SOC

**Fig. 6.** Heat maps showing relations between the SOC intervals and PCCs for SOC-based HIs.

## 2.5 Driving behavior-guided HIs evaluation

The five SOC-based HIs could be further fused according to their operating scenarios by directly extending the feature dimension, as listed in Table 3. While the first three fusion strategies required either a charge or discharge scenario for feature extraction, the fusion strategy 4 required both. The evaluation results of the five SOC-based HIs and four fusion strategies are shown in Fig. 7 (the detailed results are listed in Supplementary Table S4 and Table S5), where the empty markers represent SOC-based HIs, and the filled markers denote fusion strategies. The line between the markers covered the recommended SOC range for each feature to be extracted, based on the analysis in Fig. 6. The line style can be further subdivided into solid lines, dash lines, and dot-dash lines. The solid line implies that the entire interval is required to reach a good performance, e.g., the ICP-SOC needs the SOC interval from 20% to 31%. In contrast, some HIs can still have good performance with only a 5% interval available, which is denoted by the dash line. The dot-dash line works just as a linker between separated intervals, where the covered interval is actually not required. Here, the SOH estimation RMSE of the WOA-ELM model is used to assess the performance of HIs, due to its better accuracy than the ELM model. Specifically, the lowest RMSEs of each HI are represented by various colors, where the color bar shows that the color would change from red to green when the RMSE is reduced. The HIs are also listed at the bottom of the figure, in order from worst to best performer. Overall, the HIs after fusing have lower RMSEs than individual SOC-based HIs, which is attributed to the wider range of SOC. As displayed in the figure notes at the bottom, the Fusion 3 strategy has been improved by 0.6% RMSE from the TRE-SOC, whereas the VRE-SOC has been improved from 1.06% to 0.81% RMSE by constructing Fusion 2. When all five SOC-based HIs are fused together, the estimation RMSE can be decreased to 0.69%.

**Table 3:** Scenario-based fusion strategies.

| Fusion strategy | Scenario | Definition |
|---|---|---|
| Fusion 1 | CC charge | VRE-SOC + ICP-SOC |
| Fusion 2 | CCCV charge | VRE-SOC + ICP-SOC + CDE-SOC |
| Fusion 3 | CC discharge | TRE-SOC + VDE-SOC |
| Fusion 4 | CCCV charge and CC discharge | VRE-SOC + ICP-SOC + CDE-SOC +TRE-SOC + VDE-SOC |

Although feature fusion strategies can improve the estimation performance in both charge and discharge scenarios, the acquisition probability could thus be reduced, as the final SOC interval should fulfill the requirements of all HIs to be fused. As can be seen in Fig. 7, the HIs located around higher SOC generally have higher acquisition probability (e.g., CDE-SOC has the highest acquisition probability of 67.44%), which is also consistent with the background histogram of the probability of the intervals with 1% resolution [70]. Overall, the feature fusion in the charge scenario is not recommended, as the practicality could be largely sacrificed. For example, the acquisition probability of the Fusion 2 strategy, which is constructed from the CDE-SOC and VRE-SOC, has fallen to 11.08%. Considering that the CDE-SOC and VRE-SOC have shown relatively good performance (i.e., 1.04% and 1.06% RMSE), the Fusion 2 strategy is not necessary. Fusion in the discharge scenario might be needed since the performance of the TRE-SOC is relatively poor. However, the practical decision-making and trade-off should depend on the specific conditions, such as the minimum requirements for the SOH estimation accuracy. When all five HIs can be extracted and fused (Fusion 4 strategy), the best estimation RMSE could be reduced to 0.69%. Nevertheless, this fusion strategy has no practical reference significance because the successful acquisition is a small-probability event with a chance of less than 1%.

The present work can be improved and further developed in the future. First, the feature evaluation and model development are conducted under constant operating conditions (i.e., protocol, C-rate). Different protocols even diverse battery chemistries can be used to screen out HIs with better generalizability. Second, as PCC can only test the linear relationship, other nonlinear correlation analysis methods will be applied for better feature screening. Finally, the progress of internet-of-things and cloud platform technology enables the collection and analysis of EV user behaviors, which can provide more information for HIs evaluation. In addition, with the cloud-edge corporation, we can use the proposed framework to develop and regularly update the cloud-based SOH estimation algorithm, and then deploy it into the edge-based embedded battery management system. Hence a cloud-EV simulation platform can be built in future work, to validate the feasibility and applicability of the proposed framework.

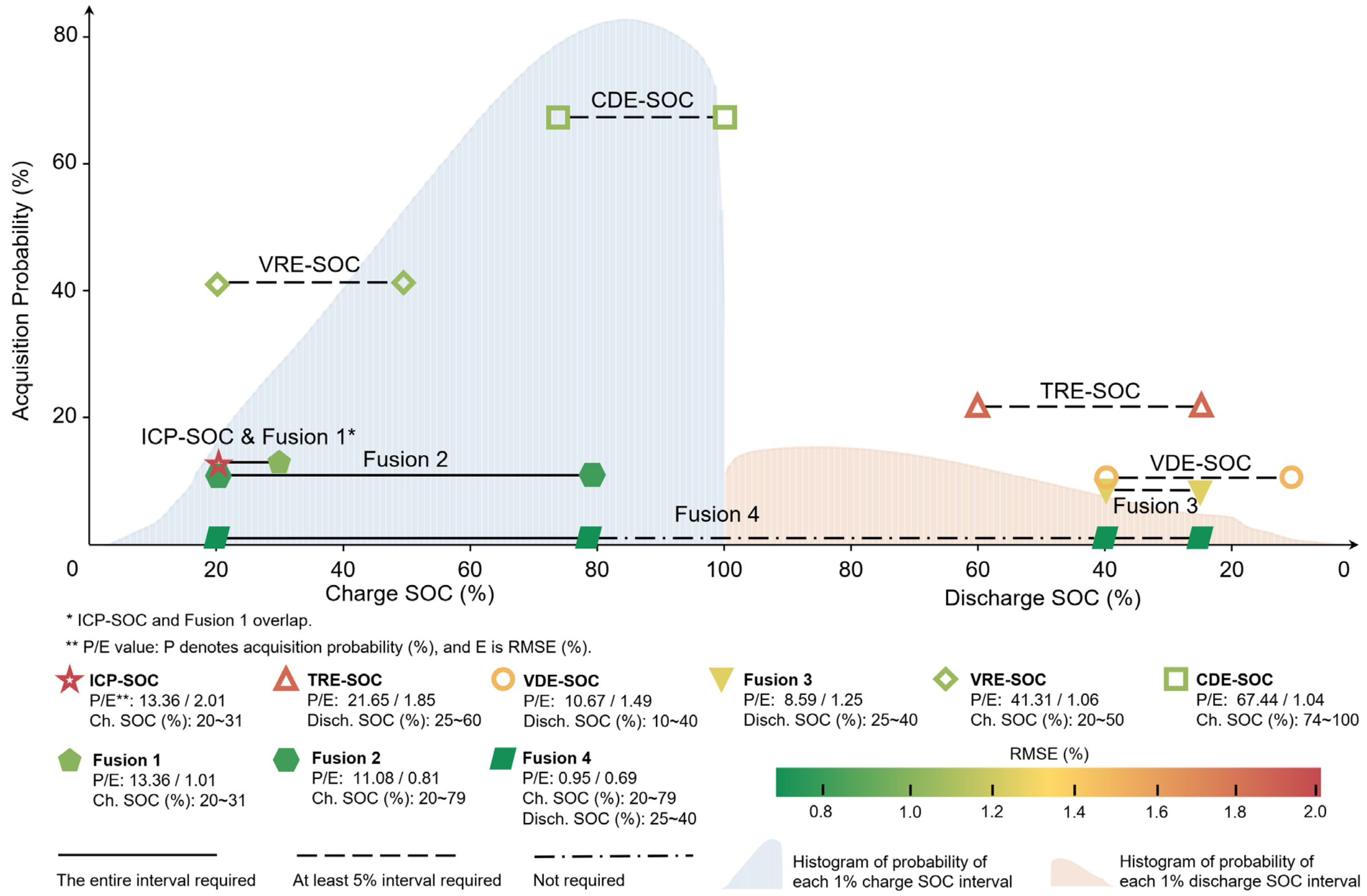

**Fig. 7.** Practicality and performance evaluation of screened five SOC-based HIs and four fusion strategies.

In summary, this work proposed a feature-based machine learning pipeline for battery health monitoring to enhance the algorithm's practicality for real-world driving behaviors, achieved by evaluating the acquisition probability of candidate HIs. To this end, the first step of the workflow would be summarizing candidate HIs and interpreting them from the perspective of degradation modes and mechanisms. Next, both the acquisition difficulty and estimation performance of candidate HIs will be assessed. Furthermore, a multi-step screening method is proposed to select superior features and avoid information redundancy, which is followed by the effects analysis of SOC intervals on HIs performance. Finally, scenario-based fusion is carried out to construct more candidate features. By introducing real-world EV driving behaviors, the feature extraction strategy can be determined by considering the acquisition probability, which can provide useful guidance. In addition, the trade-off between the performance and practicality of fusion strategies could be well balanced by tracking the latest driving behavior statistics. This work highlights the importance of the practicality evaluation of HIs for feature-based battery health monitoring algorithms.

## Credit author statement

**Nanhua Jiang**: Analysis, Software, Validation, Visualization, Writing - Original Draft.

**Jiawei Zhang**: Analysis, Validation, Visualization, Writing - Review & Editing.

**Weiran Jiang**: Analysis, Writing - Review & Editing.

**Yao Ren**: Analysis, Writing - Review & Editing.

**Jing Lin**: Analysis, Writing - Review & Editing.

**Edwin Khoo**: Analysis, Writing - Review & Editing.

**Ziyou Song**: Analysis, Conceptualization, Supervision, Writing - Review & Editing.

### Declaration of competing Interest

The authors have no conflict of interests to declare about this work.

## Acknowledgements

J.L. and E.K. acknowledge funding by Agency for Science, Technology and Research (A*STAR) under the Career Development Fund (C210112037).

# Supplementary Information

**for**

Driving behavior-guided battery health monitoring for electric vehicles using machine learning

**This Word file includes:**

- **Table S1**. Information for large-scale publicly available EV driving dataset.
- **Table S2**. Specification of three battery datasets and selected 12 cells in this work.
- **Table S3**. Results of proposed SOC-based HIs.
- **Table S4**. Results of scenario-based HI fusion.
- **Table S5**. The final selected intervals of partial-curve-based HIs.
- **Fig. S1**. Temperature profiles.
- **Fig. S2**. The frequency of various SOC intervals under real-world operations.
- **Fig. S3**. Battery performance degradation data from NASA dataset.
- **Note S1**. The specifications of the three battery degradation datasets.
- **Note S2**. The specific definitions of partial time-based HIs.
- **Note S3**. The specific definitions of four statistics-based HIs.

## Table S1. Information for large-scale publicly available EV driving dataset.

Metadata of EV operating data; each record contains multiple real-world states of vehicles during a driving/charging session.

| Battery Feature | Description | Resolution |
|---|---|---|
| b.start_soc | Start state of charge (%) | 1% |
| b.end_soc | End state of charge (%) | 1% |
| b.max_total_voltage | Max total voltage of the battery system (V) | 0.1 V |
| b.min_total_voltage | Min total voltage of the battery system (V) | 0.1 V |
| b.max_total_current | Max absolute total current of the battery system (A) | 0.1 A |
| b.min_total_current | Min absolute total current of the battery system (A) | 0.1 A |
| b.max_cell_volt | Max voltage of cells (V) | 0.001 V |
| b.min_cell_volt | Min voltage of cells (V) | 0.001 V |
| b.max_acquisition_point_temp | Max temperature of probes in the battery system (°C) | 1°C |
| b.min_acquisition_point_temp | Min temperature of probes in the battery system (°C) | 1°C |
| b.start_max_cell_volt | Max cell voltage at starting time of this segment (V) | 0.001 V |
| b.start_min_cell_volt | Min cell voltage at starting time of this segment (V) | 0.001 V |
| b.end_max_cell_volt | Max cell voltage at ending time of this segment (V) | 0.001 V |
| b.end_min_cell_volt | Max cell voltage at ending time of this segment (V) | 0.001 V |
| b.start_max_acquisition_point_temp | Max probe temperature of the battery system at starting time (°C) | 1°C |
| b.start_min_acquisition_point_temp | Min probe temperature of the battery system at starting time (°C) | 1°C |
| b.end_max_acquisition_point_temp | Max probe temperature of the battery system at ending time (°C) | 1°C |
| b.end_min_acquisition_point_temp | Min probe temperature of the battery system at ending time (°C) | 1°C |
| b.start_total_current | Total current of the battery system at starting time (A) | 0.1 A |
| b.start_total_volt | Total voltage of the battery system at starting time (V) | 0.1 V |
| b.end_total_current | Total current of the battery system at ending time (A) | 0.1 A |
| b.end_total_volt | Total voltage of the battery system at ending time (V) | 0.1 V |
| b.max_power | Max power (kW) | 1 W |
| b.min_power | Min power (kW) | 1 W |
| b.avg_current | Average current (A) | 0.1 A |
| b.avg_volt | Average voltage (V) | 0.1 V |
| b.charge_c | Charging rate (C) | 0.1 |
| b.power | Charing energy in charging segments OR energy consumption in driving segments | 0.1 kWh |
| b.volume | Cumulative capaticy (Ah) | 0.1 Ah |
| b.start_avg_temp | Average temperature of the battery at starting time (°C) | 0.01°C |
| b.end_avg_temp | Average temperature of the battery at ending time (°C) | 0.01°C |
| b.max_avg_temp | Average of max battery temperatures (°C) | 0.1°C |
| b.min_avg_temp | Average of min battery temperatures (°C) | 0.1°C |
| b.max_min_temp | Max value of min battery temperatures (°C) | 1°C |
| b.min_max_temp | Min value of max battery temperatures (°C) | 1°C |
| **Vehicle Feature** | **Description** | **Resolution** |
| b.st_time_e | Start time of a driving/charging segment (unix epoch ms) | 1–30 s |
| b.et_time_e | End time of a driving/charging segment (unix epoch ms) | 1–30 s |
| b.category | 10: driving; 20: stop; 30: charging; 60: not activated; 70: fault; 50: full charging | - |
| b.start_mileage | Start total driving distance (km) | 0.1 km |
| b.stop_mileage | End total driving distance (km) | 0.1 km |
| b.avg_speed | Average speed (km/h) | 0.1 km/h |
| b.max_speed | Max speed of this segement (km/h) | 0.1 km/h |
| b.start_longitude | Longitude of the vehicle at starting time | $10^{-5}$~$10^{-4}$ (1–10 m) |
| b.start_latitude | Latitude of the vehicle at starting time | $10^{-5}$~$10^{-4}$ (1–10 m) |
| b.end_longitude | Longitude of the vehicle at ending time | $10^{-5}$~$10^{-4}$ (1–10 m) |
| b.end_latitude | Latitude of the vehicle at ending time | $10^{-5}$~$10^{-4}$ (1–10 m) |
| b.today_start_mileage | Total driving distance (km) at the beginning of this day. | 0.1 km |
| b.today_end_mileage | Total driving distance (km) at the end of this day. | 0.1 km |
| b.avg_acquisition_point_temp | Average probe temperature (°C) | 0.01°C |
| b.avg_motor_temp | Averge motor temperature (°C) | 0.1°C |
| b.max_motor_temp | Max motor temperature (°C) | 1°C |
| b.min_motor_temp | Min motor temperature (°C) | 1°C |
| b.avg_motor_rpm | Average motor rpm (r/min) | 0.1 r/min |
| b.max_motor_rpm | Max motor rpm (r/min) | 1 r/min |
| b.min_motor_rpm | Min motor rpm (r/min) | 1 r/min |
| **Other Feature** | **Description** | **Resolution** |
| b.vid | Anonymous unique ID | - |
| b.vin | Anonymous VIN | - |
| b.start_temp | Ambient temperature at starting time (°C) | 1°C |
| b.end_temp | Ambient temperature at ending time (°C) | 1°C |

**Table S2. Specification of three battery datasets and selected 12 cells in this work.**

| Dataset | NASA | CALAE | MIT |
|---|---|---|---|
| Manufacturer | LG | - | A123 Systems |
| Form factor | 18650 Cylindrical | Prismatic | 18650 Cylindrical |
| Cell anode | Graphite | Graphite | Graphite |
| Cell cathode | LiNiCoAlO2 | LiCoO2 | LiFePO4 |
| Charge conditions | CC-CV | CC-CV | Multi-step fast-charging |
| Charge rate | 0.75C | 0.5C | 4.8C/1C |
| Discharge rate | 1C | 1C | 4C |
| Upper cut-off voltage | 4.2 V | 4.2 V | 3.6 V |
| Lower cut-off voltage | 2.7 V, 2.5 V, 2.2 V, 2.5 V | 2.7 V | 2.0 V |
| Ambient temperature | 24°C | 24°C | 30°C |
| Nominal capacity | 2 Ah | 1.1 Ah | 1.1 Ah |
| Number of cells | 4 | 4 | 4 |
| Cell number | B5, B6, B7, B18 | B35, B36, B37, B38 | B6, B9, B17, B25 |

**Table S3. Results of proposed SOC-based HIs.**

| Health indicators | Mode | PCC | ELM | ELM-WOA | SOC interval | Probability |
|---|---|---|---|---|---|---|
| VRE-SOC | CC charge | -0.9937 | 1.08% | 1.06% | 20-50%[5%] | 41.31% |
| ICP-SOC | CC charge | 0.9888 | 2.11% | 2.01% | 20-31% | 13.36% |
| CDE-SOC | CV charge | 0.9892 | 1.12% | 1.04% | 74-100%[5%] | 67.44% |
| TRE-SOC | CC discharge | -0.9888 | 1.86% | 1.85% | 25-60%[5%] | 21.65% |
| VDE-SOC | CC discharge | -0.9935 | 1.49% | 1.49% | 10-40%[5%] | 10.67% |

**Table S4. Results of scenario-based HI fusion.**

| Case | HI fusion | Scenario | ELM | ELM-WOA | Probability |
|---|---|---|---|---|---|
| 1 | VRE-SOC<br>ICP-SOC | CC charge | 3.85% | 1.01% | 13.36% |
| 2 | VRE-SOC<br>ICP-SOC<br>CDE-SOC | CC-CV charge | 2.43% | 0.81% | 11.08% |
| 3 | TRE-SOC<br>VDE-SOC | CC discharge | 1.40% | 1.25% | 8.59% |
| 4 | VRE-SOC<br>ICP-SOC<br>CDE-SOC<br>TRE-SOC<br>VDE-SOC | Charge and discharge | 0.87% | 0.69% | 0.95% |

**Table S5. The final selected intervals of partial-curve-based HIs with grid search.**

| | HIs | NASA | CALCE | MIT |
|---|---|---|---|---|
| CC charge Voltage | TEVR | 3.9V-4V | 3.7V-4.2V | 3.48V-3.59V |
| | VRET | 925s-975s | 3000s-3500s | 200s-400s |
| | SCV | 3.9V-4V | 3.7V-4.2V | 200s-400s |
| CV charge Current | TECD | 0.5A-1.2A | 0.05A-0.3A | 0.25A-1A |
| | CDET | 3700s-4800s | 200s-800s | 650s-700s |
| | SCC | 0.5A-1.2A | 0.05A-0.3A | 0.25A-1A |
| CC discharge Voltage | TEVD | 3.4V-3.8V | 3.4V-4.1V | 2.1V-3.5V |
| | VDET | 0s-1000s | 50s-1500s | 100s-650s |
| | SDV | 3.4V-3.8V | 3.4V-4.1V | 2.1V-3.5V |
| CC discharge Temperature | TETR | 29℃-36℃ | / | / |
| | TRET | 1000s-1400s | / | / |
| | SDT | 29℃-36℃ | / | / |

As shown in **Table S5**, the optimal intervals for partial-curve-based HIs are listed in this table, where the grid search method is used for optimizing the interval. The initial step is to determine the boundary for searching. For voltage-relevant HIs, the range for searching is from the upper cut-off voltage to the lower cut-off voltage. For temperature-relevant HIs, the range for searching is from the highest temperature to the lowest temperature. The available current intervals range from 0A to the maximum current. For the time-relevant HIs, the boundary for searching depends on the specific operating time. The performance of all combinations is then examined using the PCC after selecting 10 points that are equally spaced within the selection range. It might be needed to repeatedly upgrade the screening resolution if the performance is still unsatisfactory.

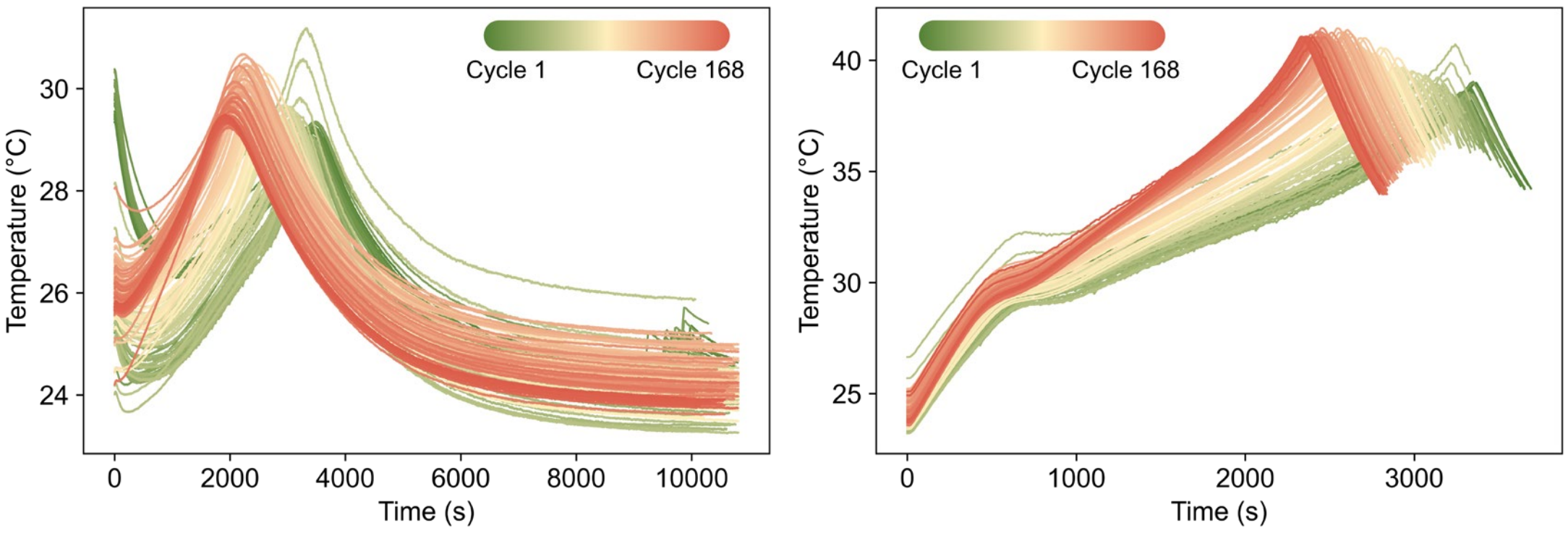

(a) (b)

**Fig. S1. Temperature profiles during (a) CC-CV charge and (b) CC discharge.**

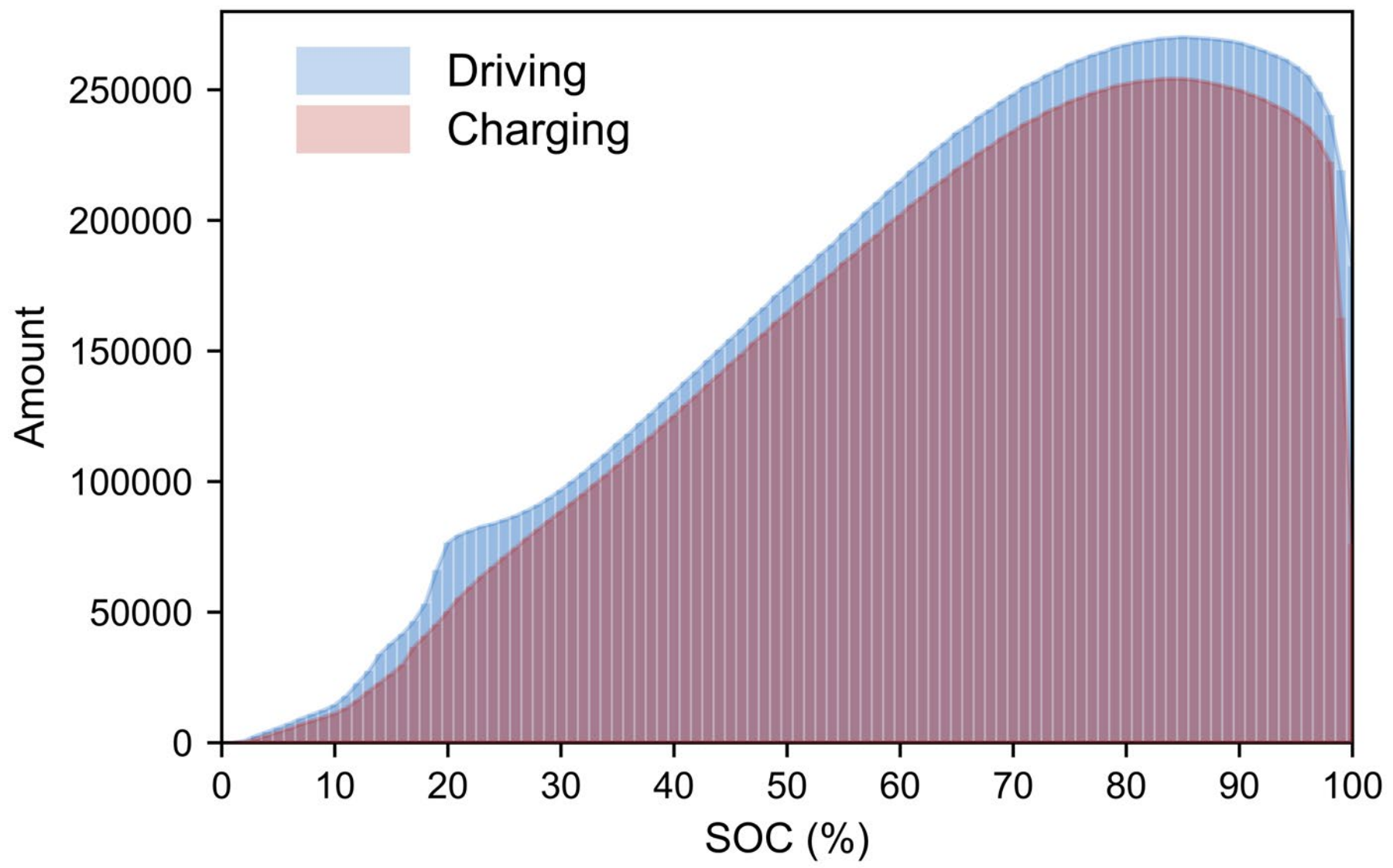

**Fig. S2. The frequency of various SOC intervals under real-world operations, with consideration of all EV data.**

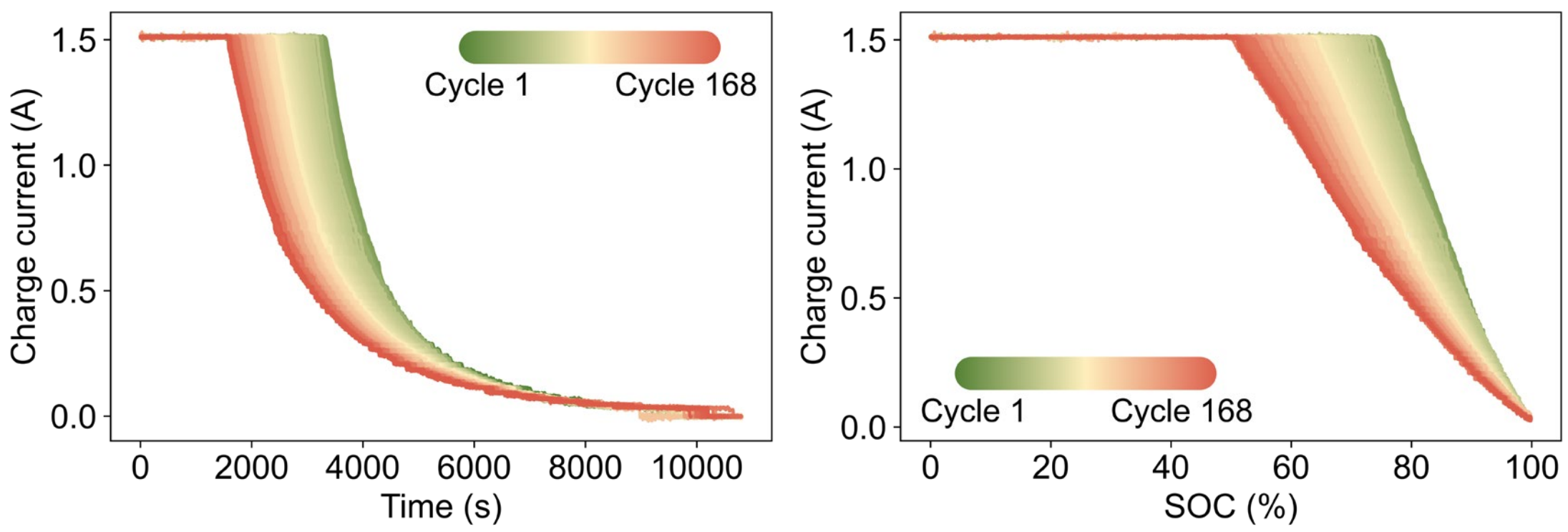

(a) (b)

**Fig. S3. Battery performance degradation data from NASA dataset (a) Time-based HI and (b) SOC-based HI.**

## Note S1. The specifications of the three battery degradation datasets.

The first dataset was obtained from the data repository of the NASA Ames Prognostics Center of Excellence [71]. This dataset was collected from commercially available 18650-size lithium-ion batteries with a rated capacity of 2Ah. While the chemistry of these 2Ah battery cells was not mentioned in [72], related work [73] presented that the tested second-generation 18650-size 1Ah lithium-ion cells are composed of lithium nickel cobalt aluminum oxide (NCA) and graphite. One commonly studied group of batteries (labeled as No. 5, 6, 7, and 18), that was cycled with 0.75C CC-CV protocol at 24℃ until 70% SOH, is used in this work. The second dataset is provided by the Center for Advanced Life Cycle Engineering (CALCE) of the University of Maryland [74]. Four lithium cobalt oxide (LCO) 1.1Ah battery cells (labeled as CS2_35, CS2_36, CS2_37, and CS2_38), which were cycled with a 0.5C CC-CV charge protocol, are selected to construct the dataset in this work. The third dataset from the Massachusetts Institute of Technology (MIT) and Stanford University contains 124 commercial 18650-size lithium iron phosphate (LFP)/graphite cells, that were cycled at a constant temperature of 30℃ with various fast-charging protocols and a fixed CC-CV discharge protocol at 4C to 2.0V with a cut-off current of C/50 [75]. The battery cells (labeled as No. 6, 9, 17, and 25) cycled with the 4.8C-1C two-step charge protocol are used in this work. With these datasets, we use discharge capacity to calculate SOH as the ground truth for performance evaluation. Whereas the discharge capacity for each cycle is provided in the NASA and MIT dataset, it is calculated by coulomb counting for the CALCE dataset.

## Note S2. The definitions of partial time-based HIs.

By counting the charge or discharge time within a certain voltage range, the TEVR and TEVD can be extracted as HIs [76]. Similarly, the time of equal current drop (TECD) [76] during CV charge and time of equal temperature rise (TETR) [77] during CC discharge can be obtained, when certain current or temperature ranges are included in the operating scenarios. Besides, similar features can be extracted by fixing the time interval as constraints and calculating the increments of measurements (i.e., current, voltage, and temperature), such as the voltage rise of equal time (VRET) [78], the voltage drop of equal time (VDET) [79], the current drop of equal time (CDET) [80], as well as the temperature rise of equal time (TRET) [81]. Note that these HIs are essentially the average slopes or their reciprocal (scaled by the window size) within certain segments of the raw time series of voltage, current, and temperature, so they fall into the same category as the HIs of explicit straight-line slope of external measurements within a certain time range, including the slope of charge current (SCC) [71, 82, 83], the slope of charge voltage (SCV) [71, 81, 82], the slope of discharge voltage (SDV) [84, 85], and the slope of discharge temperature (SDT) [86].

## Note S3. The specific definitions of four statistics-based HIs.

Note that extracting statistics from a time series is essentially the same as integrating a certain function of the variable $X$ over a time interval:

$$E_{[t_1,t_2]}[f(X)] = \frac{1}{t_2 - t_1}\int_{t_1}^{t_2} f\big(X(t)\big)dt \approx \frac{\sum_i f(X_i)\Delta t}{t_2 - t_1} = \frac{1}{n}\sum_i f(X_i), \qquad (6)$$

where the $X_i$'s with $i = 1, \dots, n$ are the time-discrete signals, so such statistics can be viewed as a special kind of integral-based HIs. Some obvious statistical HIs include those based on the first four moments of $X(t)$, i.e. mean, variance, skewness, and kurtosis:

$$\mu = \frac{1}{n}\sum_{i=1}^{n} X_i, \qquad SD = \sqrt{\frac{1}{n}\sum_{i=1}^{n}(X_i - \mu)^2}, \qquad SK = \frac{1}{n\sigma^3}\sum_{i=1}^{n}(X_i - \mu)^3, \qquad KT = \frac{1}{n\sigma^4}\sum_{i=1}^{n}(X_i - \mu)^4. \qquad (7)$$

Standard deviation (SD), which is the square root of variance, was used as an HI by [87]. Skewness (SK) measures the extent to which the probability distribution of the signal leans toward one side of its mean, which was used by [87, 88]. Kurtosis (KT) measures the "peakedness" of the probability distribution of the signal, which was used by [87, 89]. In principle, these statistics can be obtained from any time interval of any variable, but in this work, we only use SD, SK and KT extracted from the full discharge voltage curve.

One drawback of the above conventional statistics is that they drop all the time-order information and are invariant to shuffling the time series, while such order may contain important information about the underlying process. Sample entropy (SE) [90] is one such statistic that take the time order into account and has been used as an HI by [91], which assesses the predictability and regularity of a time series by estimating the conditional probability that two randomly chosen sub-series are close to each other at the $(m + 1)$-th position, given that they are close in their first $m$ positions. In particular, the $m$-th order SE can be calculated for a length-$n$ ($n \geq m$) segment of $X_i$'s as follows. Denote the length-$m$ window starting at $i$ by

$$X_i^{(m)} = [X_i, X_{i+1}, \dots, X_{i+m-1}], \qquad i = 1, \dots, n - m + 1. \qquad (8)$$

The maximum norm

$$\| X_i^{(m)} - X_j^{(m)} \| = \max_{0 \leq k \leq m-1} |X_{i+k} - X_{j+k}| \qquad (9)$$

is used to quantify the distance between two windows of the same length, so by two windows $X_i^{(m)}$ and $X_j^{(m)}$ being close, we mean $\| X_i^{(m)} - X_j^{(m)} \| \leq r$ for some specified tolerance $r > 0$. Now the probability of two randomly chosen sub-series being close at the first $m$ positions can be estimated as

$$A^m(r) = \frac{\#(i,j)\text{pairs: } \left\| X_i^{(m)} - X_j^{(m)} \right\| \leq r, 1 \leq i < j \leq n - m}{(n - m)(n - m - 1)}, \qquad (10)$$

where we drop the last length-$m$ window to match the windows used for the length-$(m + 1)$ calculation:

$$B^m(r) = \frac{\#(i,j)\text{pairs: } \left\| X_i^{(m+1)} - X_j^{(m+1)} \right\| \leq r, 1 \leq i < j \leq n - m}{(n - m)(n - m - 1)}. \qquad (11)$$

Finally, the SE of order $m$ with tolerance $r$ estimated from a length-$n$ series can be calculated as

$$SE(m, r, n) = -ln\left[\frac{B^m(r)}{A^m(r)}\right], \qquad (12)$$

where $m$ and $r$ are set as 1 and 0.15 for the SE used as an HI in this work.

## References for Supplementary Information